\documentclass[10pt]{article}

\usepackage{amsmath}
\usepackage{amsthm}
\usepackage{amssymb}

\chardef\bslash=`\\ 

\makeatletter
\def\verbatim{\interlinepenalty\@M \@verbatim
  \leftskip\@totalleftmargin\advance\leftskip2pc
  \frenchspacing\@vobeyspaces \@xverbatim}
\makeatother
\theoremstyle{plain}

\theoremstyle{remark}

\numberwithin{equation}{section}

\def\1I{\relax{\rm 1\kern-.25em \rm l}} 

\newcommand{\unity}{\1I}

\renewcommand{\thefootnote}{${\Large *}_{\arabic{footnote}}$}

\def\Rahmen#1#2#3 {
   \vbox{\hrule height#2
         \hbox{
               \vrule width#2
               \hskip#1
               \vbox{
                     \vskip#1{}
                     \hbox{#3}
                     \vskip#1
                    }%
               \hskip#1
               \vrule width#2
              }
         \hrule height#2
        }}
\def\href#1#2{#2}

\begin{document}

\thispagestyle{empty}
\rightline{hep-th/0303085}
\vspace{2truecm}
\centerline{\bf \Large A Note on Exotic Fractions of Supersymmetries}
\vspace{.5truecm}

\newcounter{Institut}
\vspace{1.5truecm}
\centerline{\bf Andr\'e Miemiec\refstepcounter{Institut}\label{Inst_Andre}{}$^{*_{\theInstitut}}$
                and 
                Igor Schnakenburg\refstepcounter{Institut}\label{Inst_Igor}{}$^{*_{\theInstitut}}$
           }

\vskip1cm
\parbox{.9\textwidth}{
\parbox{.4\textwidth}{
\centerline{$^{*_{\ref{Inst_Andre}}}$~ Department of Theoretical Physics}
\centerline{                           University of Turin}
\centerline{                           10125 Torino}
\centerline{                           Italia}
\centerline{                           miemiec@to.infn.it}
}\hfill
\parbox{.4\textwidth}{
\centerline{$^{*_{\ref{Inst_Igor}}}$~  Racah Institute of Physics}
\centerline{                           The Hebrew University} 
\centerline{                           Jerusalem 91904} 
\centerline{                           Israel}
\centerline{                           igorsc@phys.huji.ac.il}
}
}

\vspace{.4truecm}

\vspace{1.0truecm}
\begin{abstract}
 \noindent
 In this note we study supersymmetric solutions of 11 dimensional 
 SUGRA. Amongst others, we find a new solution which preserves 3/8 
 of the original supersymmetry. 
\end{abstract}
\bigskip \bigskip
\newpage


\noindent
\section{Introduction}

Solutions of supergravity theories which preserve 1/2 of the original 
supersymmetry typically occur in the description of brane solutions.
Using the techniques of intersecting brane solutions one can reduce 
this number quite generically to any fraction 1/$2^n$, with $n$ 
an integer. Other fractions are less simple to understand and are called 
exotic.
In \cite{Gauntlett:2002nw} the authors discovered a new solution of 
 11 dimensional supergravity, which preserved a fraction of $\nu\,=\,5/8$ 
supersymmetries. In a following paper \cite{Gauntlett:2002fz} it was 
shown that this solution fits into a whole class of solutions and the 
questions was raised if further solutions with exotic fractions of 
preserved supersymmetry can be found within this class. This note 
confirms in a case by case study that this actually is the case\footnote
{
  After finishing our work, we became aware of the nice work 
  of \cite{Harmark:2003ud}. In this paper plenty of similar 
  solutions with exotic supersymmetry are constructed. 
}.

\noindent
The class of solutions proposed in \cite{Gauntlett:2002fz} are 
\begin{eqnarray}
   ds_{11}^2 &=& -\,(\,dt\,+\,\omega\,)^2 ~+~ ds^2({\mathbb{R}}^{10})
                 \nonumber\\
           F &=& -d\omega\wedge\Omega
\end{eqnarray}
with $\Omega$ defined in terms of complex coordinates 
$z^a\,=\,x^{2a-1}+i\,x^{2a}$ of 
$~{\mathbb{C}}^{5}\,=\,{\mathbb{R}}^{10}$,
\begin{eqnarray}
  \Omega &=& \frac{i}{2}\,\sum\limits_{a=1}^5 dz^a\wedge d{\bar{z}}^a~,
\end{eqnarray}
and $d\omega\,=\,\alpha\,+\,\bar{\alpha}\,\subset\,\Lambda^{2,0}({\mathbb{C}}^{5})\oplus\Lambda^{0,2}({\mathbb{C}}^{5})$.\\

\noindent
The Einstein equation and the equation of motion for the 4-form field 
strength read
\begin{eqnarray}
     R_{ab}\,-\,\frac{1}{12}\,
                 \left(\,
                            F_{ac_1c_2c_3}F_{b}{}^{c_1c_2c_3}
                            \,-\,\frac{1}{12}\,\eta_{ab}\,
                             F_{c_1c_2c_3c_4}F^{c_1c_2c_3c_4}\,
                 \right) &=& 0\label{Einstein}~,\\
     d(\ast F) ~+~ \frac{1}{2}\,F\wedge F &=&0\label{4Form}~,
\end{eqnarray} 
while the Killing spinor equation is given by
\begin{eqnarray}\label{KSE}
     \partial_\mu\varepsilon\,-\,
     \frac{1}{4}\omega_{\mu ab}\Gamma^{ab}\,\varepsilon ~+~ \frac{1}{288}
     \left(\, 
              \Gamma_{\mu}\Gamma^{\nu_1\nu_2\nu_3\nu_4} ~-~ 12\,
              \delta_{\mu}^{\nu_1}\Gamma^{\nu_2\nu_3\nu_4}\,
     \right)\,F_{\nu_1\nu_2\nu_3\nu_4}\varepsilon &=& 0~.
\end{eqnarray}

\section{Examples}

\subsection{First Example}

\noindent
The holomorphic two form $\alpha$ we have chosen is 
\begin{eqnarray}
   \alpha &=&\frac{\gamma}{2}\,
                \left(\, 
                         dz^1\wedge dz^2 ~+~ dz^3\wedge dz^4\,
                \right)
\end{eqnarray}
which leads to 
\begin{eqnarray}
     \omega&=&\frac{\gamma}{2}\,
                \left(\,
                         -x^3dx^1+x^1dx^3-x^2dx^4+x^4dx^2
                         -y^3dy^1+y^1dy^3-y^2dy^4+y^4dy^2\,
                \right)
\end{eqnarray}

\noindent
The equations of motion (\ref{Einstein}) and (\ref{4Form}) are both 
satisfied. The Ricci tensor in a tangent frame is diagonal and given 
by $R_{ab}\,=\,\gamma^2\cdot{\rm diag}(2,1/2,\ldots,1/2,0,0)$ 
while the contractions of the field strength are:
\begin{eqnarray}
    F_{c_1c_2c_3c_4}F^{c_1c_2c_3c_4} &=& 288\,\gamma^2\\
    F_{ac_1c_2c_3}F_{b}{}^{c_1c_2c_3}&=& \left\{
                                                \begin{array}{cl}
                                                  0  & (a,b)=(0,0)\\
                                                 30\,\gamma^2 
                                                     & (a,b)=(1,1),\ldots
                                                             ,(8,8)\\
                                                 24\,\gamma^2 
                                                     & (a,b)=(9,9),
                                                       (10,10)\\
                                                  0  & a\neq b
                                                \end{array}
                                         \right.
\end{eqnarray}
Obviously eq.~(\ref{Einstein}) is satisfied. 
The second term in the equation of motion for the 4-form field strength 
(\ref{4Form}) reduces to 
\begin{eqnarray*}
 \frac{1}{2}\,F\wedge F &=& 2\,\gamma^2\,\left(\,2\,dx^{12345678}
                            +dx^{1234569 1\hspace{-1.75pt}0}
                            +dx^{1234789 1\hspace{-1.75pt}0}
                            +dx^{1256789 1\hspace{-1.75pt}0}
                            +dx^{3456789 1\hspace{-1.75pt}0}\,\right)
\end{eqnarray*}
and comparing this with $d(\ast F)$ one finds that both add up to zero.\\

\noindent
The Killing spinor equation (\ref{KSE}) can be written more 
symbolically as
\begin{eqnarray}
     \partial_\mu\varepsilon\,-\,{\mathbb{M}}_\mu\varepsilon &=& 0~.
\end{eqnarray}
By just solving this equation for the present ansatz one obtains 
12 constant Killing spinors. These 12 solutions exhaust the whole set 
of solutions. 
Now we give the argument from which we draw this conclusion.\\

\noindent
As a preparation we would like to shed some light onto the terms which we
gathered in
\begin{eqnarray}
  e_c{}^{\mu}{\mathbb{M}}_\mu &=& \frac{1}{4}\underbrace{\omega_{cab}
       \Gamma^{ab}}_{(3)}
                        ~-~ \frac{1}{288}
                        \left(\vbox{\vspace{2ex}}\right.\, 
                            \Gamma_{c}\underbrace{
                            (\Gamma^{\nu_1\nu_2\nu_3\nu_4}
                          F_{\nu_1\nu_2\nu_3\nu_4})}_{(1)} ~-~ 12\,e_c{}^{\mu}
                            \underbrace{
                              \Gamma^{\nu_2\nu_3\nu_4}F_{\mu\nu_2\nu_3\nu_4}
                            }_{(2)}
                        \left.\vbox{\vspace{2ex}}\right)\,
\end{eqnarray}
The components of the field strength $F_{\nu_1\nu_2\nu_3\nu_4}$ are 
constant and those involving the time direction vanish. Also, Gamma matrices 
with upper spatial indices are identical to the tangent frame 
Gamma matrices.  We conclude that the expressions (1) and (2) are constant. 
The spin connection in (3) turns out to be constant too. So the only source of 
coordinate dependence comes from contraction with the vielbein in the last 
term. 
Since the field strength components which include a time direction vanish, 
the only remaining source for a coordinate dependence is set to zero.\\

\noindent
This example gives a simple necessary condition, which each solution 
must satisfy
\begin{eqnarray}\label{necessary}
      0 &=& [\,\partial_\mu,\,\partial_\nu\,]\,\varepsilon 
        ~=~ e_{\mu}{}^ae_{\nu}{}^b\,
            \left(\,  2\,\omega_{[ab]}{}^c\,{\mathbb{M}}_c ~+~
                      [\,{\mathbb{M}}_b,\,{\mathbb{M}}_a\,]\,
            \right)\,\varepsilon~,
\end{eqnarray}
where  ${\mathbb{M}}_a\,=\,e_a{}^\mu{\mathbb{M}}_\mu$ and
$\omega_{[\mu\nu]}{}^a\,=\,\partial_{[\mu}e_{\nu]}{}^a$. In deriving 
eq.~(\ref{necessary}) we used the fact that in the present case all 
${\mathbb{M}}_a$ are constant. Since the only nonvanishing components 
of $\omega_{[\mu\nu]}{}^a$ are
\begin{align}
    \omega_{[13]}{}^0 &=\gamma~, & 
    \omega_{[24]}{}^0 &=-\gamma~, &
    \omega_{[57]}{}^0 &=\gamma~, &
    \omega_{[68]}{}^0 &=-\gamma& 
\end{align}
equation (\ref{necessary}) evaluated for $(a,b)=(0,1)$ just reduces
to 
\begin{eqnarray}
     0 &=& [\,{\mathbb{M}}_1,\,{\mathbb{M}}_0\,]\,\varepsilon~, 
\end{eqnarray}
the kernel of which is just the span of the 12 Killing spinors mentioned 
above. So we already found all solutions.

\subsection{Second Example}

We now take the following ansatz for 
\begin{eqnarray}
   \alpha &=& \frac{\gamma}{2}\,
              \left(\,
                  dz^1\wedge dz^2 ~+~ dz^1\wedge dz^3 ~+~ dz^2\wedge dz^3\,
              \right)~.
\end{eqnarray}
In this case the Ricci tensor in the tangent frame is no longer diagonal. 
Nevertheless the equations of motion are satisfied both for the metric 
and the four form field strength.  
One finds 16 constant Killing spinors $\varepsilon_i$ generating the 
16 dimensional kernels of ${\mathbb{M}}_1$ to ${\mathbb{M}}_6$. 
They are contained in the 20 dimensional kernels of ${\mathbb{M}}_0$ and 
${\mathbb{M}}_7$ to ${\mathbb{M}}_{10}$. The form of 
the Killing spinors can be calculated explicitly and is given for 
convenience in appendix \ref{AppendixB}. The kernels of ${\mathbb{M}}_0$ and 
${\mathbb{M}}_7$ to ${\mathbb{M}}_{10}$ are all equal and the additional 
four dimensions are spanned by $\vartheta$, which satisfies
\begin{eqnarray*}
     0 ~\neq ~ {\mathbb{M}}_i(\vartheta)  ~\subset~ 
       {\rm span}(\,\varepsilon_1,\ldots \varepsilon_{16}\,)~,
      \quad\quad\quad i~=~1,\ldots , 6
\end{eqnarray*}
and so we can apply the same trick used in \cite{Gauntlett:2002nw} to
construct a further Killing spinor by setting
\begin{eqnarray*}
        \varepsilon &=& \vartheta ~+~ x^i\,{\mathbb{M}}_i\,\vartheta~.
\end{eqnarray*}
Finally we obtain 20 Killing spinors, i.e. $\nu\,=\,5/8$ of the original supersymmetries are preserved.\\

\section{Conclusion}

In the above we have presented two solutions, which preserve two different 
fractions of supersymmetry. The first example preserves $3/8$, the later 
$5/8$. The first example preserves a different exotic fraction of 
supersymmetry than the example already discovered in \cite{Gauntlett:2002fz}.
We have also tried different ans\"atze for $\alpha$. Their precise form and 
the amount of supersymmetry they preserve can be found in the table below.
\begin{center}
\begin{tabular}{|c|l|}
\hline
                       &\\[-1.5ex] 
   $\alpha$            &   $\nu$\\[.5ex]
\hline
                       &\\[-1ex] 
   $dz^1\wedge dz^2$   &    5/8$^{\ast_2}$   \\
   $dz^1\wedge dz^2\,+\,dz^1\wedge dz^3$ & 5/8\\
   $dz^1\wedge dz^2\,+\,dz^1\wedge dz^3\,+\,dz^2\wedge dz^3$ & 5/8\\
   $dz^1\wedge dz^2\,+\,dz^3\wedge dz^4$ & 3/8\\
   $dz^1\wedge dz^2\,+\,dz^3\wedge dz^4\,+\,dz^1\wedge dz^5$ & 3/8\\
   $dz^1\wedge dz^2\,+\,dz^2\wedge dz^3\,+\ldots+\,dz^5\wedge dz^1$ & 3/8\\[.5ex]
\hline
\end{tabular}\\
\end{center}
%
\renewcommand{\thefootnote}{}
\footnote{$^{\ast_2}$~This is actually the case considered in \cite{Gauntlett:2002nw}}
\renewcommand{\thefootnote}{${\Large *}_{\arabic{footnote}}$}
%
In particular, we have only found solutions which preserve either $3/8$ 
or $5/8$ of the original supersymmetry. It would be interesting to see 
whether this ansatz could also produce solutions with other numbers of 
preserved supersymmetries.

\vskip1cm

\begin{appendix}

\section{Clifford Algebra}

The 11 dimensional Clifford Algebra with signature 
$\eta^{ab}\,=\,{\rm diag} (-1,1,\ldots, 1)$ satisfies
\begin{eqnarray}
    \{\,\Gamma^a,\,\Gamma^b\,\} &=& 2\,\eta^{ab}\,\unity_{32}~.
\end{eqnarray} 
A representation by real matrices $\Gamma^a$ can be obtained by 
taking appropriate tensor products of the Pauli matrices 
\begin{align*}
  \tau_1 &= \left(\begin{array}{cc}
                     0 & 1\\
                     1 & 0\\
                   \end{array}\right) &
  \tau_2 &= \left(\begin{array}{cc}
                     0 & -i\\
                     i & 0\\
                   \end{array}\right) &
  \tau_3 &= \left(\begin{array}{cc}
                     1 & 0\\
                     0 & -1\\
                   \end{array}\right)~. &
\end{align*}
A particular choice is ($\epsilon=i\tau_2$):
\begin{align*}
   \Gamma^0 &= -\epsilon\otimes\unity\otimes\unity\otimes\unity\otimes\tau_3 &
   \Gamma^1 &= \epsilon\otimes\epsilon\otimes\epsilon\otimes\epsilon\otimes\tau_1\\  
\Gamma^2 &= \epsilon\otimes \unity\otimes\tau_1\otimes\epsilon\otimes\tau_1 &
\Gamma^3 &= \epsilon\otimes \unity\otimes\tau_3\otimes\epsilon\otimes\tau_1\\
\Gamma^4 &= \epsilon\otimes \tau_1\otimes\epsilon\otimes\unity\otimes\tau_1 &
\Gamma^5 &= \epsilon\otimes \tau_3\otimes\epsilon\otimes\unity\otimes\tau_1\\
\Gamma^6 &= \epsilon\otimes \epsilon\otimes\unity\otimes\tau_1\otimes\tau_1 &
\Gamma^7 &= \epsilon\otimes \epsilon\otimes\unity\otimes\tau_3\otimes\tau_1\\
\Gamma^8 &= \epsilon\otimes \unity\otimes\unity\otimes\unity\otimes\epsilon &
   \Gamma^9 &= \tau_1\otimes\unity_{16}\\
   \Gamma^{10} &= \tau_3\otimes\unity_{16}
\end{align*}

\section{Explicite Killing spinors}
\label{AppendixB}

\begin{eqnarray*}
   \varepsilon &=& (\,a_1+a_2+a_3, a_4, a_5, a_6, a_7, a_8, -a_2, -a_4, 
                      a_9, a_6, a_2+a_3+a_{10}, a_4, a_{11}, a_{12}, 
                      a_5+a_{13}+a_{14},\\
                && ~\;a_6, a_5-a_9+a_{13}, -a_6, a_{10}, a_4, a_{15}, -a_{12}, 
                    a_{14}, a_6, a_1, a_4, a_{13}, -a_6, a_{16}, a_8, -a_3, 
                    a_4\,)^T~\\[1ex]
   \vartheta &=& (\,0, a_{17}, 0, a_{18}, 0, a_{19}, 0, -a_{17}, 0, a_{18}, 
                   0, a_{17}, 0, a_{20}, 0, a_{18}, 0, a_{18}, \\
             && ~\;0, -a_{17},0, a_{20}, 0, -a_{18}, 0, -a_{17}, 0, a_{18}, 0, -a_{19}, 0, 
                   -a_{17}\,)^T
\end{eqnarray*}
\end{appendix}


\end{document}